\documentclass[5p,times]{elsarticle}
  \usepackage[utf8]{inputenc}
  \usepackage{setspace}
  \usepackage{graphicx}
  \usepackage{siunitx}

\usepackage{lineno,hyperref}
\modulolinenumbers[5]

\journal{Nuclear Instruments and Methods in Physics Research Section A}

\begin{document}

\begin{frontmatter}

\title{On the use of NaI scintillation for high stability nuclear decay rate measurements}
\author{ Scott D. Bergeson \& Michael J. Ware}
\address{BYU Physics and Astronomy, Provo, UT 84602}

\author{ Jeremy Hawk }
\address{ Utah Valley Regional Medical Center, Provo, UT 84602 }

\begin{abstract}
We demonstrate the linearity and stability of a gamma-ray scintillation
detector comprised of a NaI(Tl) crystal and a scientific-grade CMOS camera.
After calibration, this detector exhibits excellent linearity more than
three decades of activity levels. Because the detector is not counting
pulses, no dead-time correction is required. When high activity sources are
brought into close proximity to the NaI crystal, several minutes are
required for the scintillation to achieve a steady state. On longer time
scales, we measure drifts of a few percent over several days. These
instabilities have important implications for precision determinations of
nuclear decay rate stability.
\end{abstract}

\begin{keyword}Gamma-ray detector \sep
    CMOS image sensors \sep
    NaI(Tl) scintillator
\end{keyword}

\end{frontmatter}

\section{Introduction}

High quality nuclear decay rate measurements require detectors with good
sensitivity, stability, and linearity. Depending on the measurement
objectives and experimental environment, the detector may also need to
provide energy resolution, timing information, and absolute activity levels.
Under ideal conditions, activity measurements are limited by Poisson
statistics. Generally speaking, the statistical limits allow improved
accuracy for higher activity samples or for longer measurement times.
However, it sometimes happens that high activity and long measurement times
compromise the sensitivity, stability, or linearity of the detector.

Improving the stability of nuclear activity measurements is particularly
important for determining variations in nuclear decay rates
\cite{Jenkins200942,Nahle20158}. It has been suggested that some beta-decay
processes are linked to solar activity. The state-of-the-art limits on
measurements of this solar-dependence is 0.01\% or perhaps slightly lower
\cite{Pomme2016281,Bergeson2017}, but the accuracy of these limits is
determined completely by detector instability \cite{Schrader2016202}. Many
decay processes result in the emission of gamma radiation.  Thus, improving
the stability of gamma-radiation detectors will help in determining the
amplitude of a seasonally dependence in nuclear decay rate.

Most gamma-ray measurements can be divided into two processes. The first
converts the gamma radiation into an electronic or photonic
signal---ionization in a gas chamber, electron-hole production in a solid,
scintillation in a solid, gas or liquid, and so forth. The second measurement
process detects and records this electronic or photonic signal, sometimes
with additional signal conditioning. High reliability measurements require
both of these processes to operate flawlessly.

These processes sometimes display systematic dependencies on the measurement
environment or configuration \cite{Pomme2016281, ware2015}. For example, the
measured count rate from a Geiger-M\"{u}ller tube depends exponentially on
the operating voltage. It also depends on the ambient pressure. As another
example, scintillation detectors using photomultiplier tubes are unable to
reliably measure high activity samples because of excessive dead-time
correction. Limitations like these determine the kinds of measurements that
these detectors can make.

In this paper we demonstrate some characteristics and limitations of a gamma
radiation detector consisting of a NaI(Tl) crystal and a scientific grade
camera. We show that the camera has good stability and acceptable linearity.
We also show nonlinearities and instabilities in NaI scintillation at high
activity levels. The regime in which we work uses higher activity samples
than can be used in traditional scintillation measurements due to dead-time
issues. This is made possible by the single-photon sensitivity and
multi-channel nature of the camera.

\section{Experimental details}

\begin{figure}
\centerline{\includegraphics{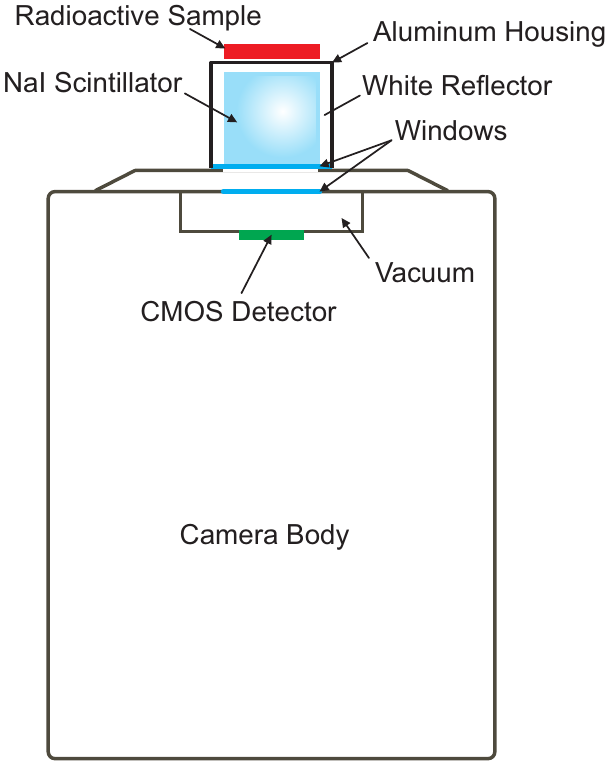}} \caption{\label{fig:schematic}
Schematic of a typical experimental setup. A radioactive sample is placed
very near the aluminum housing of a NaI scintillation crystal.  The crystal
is mounted in a light-tight manner to the top of the camera, directly above
the CMOS detector.}
\end{figure}

The equipment consists of a scientific camera, a NaI(Tl) scintillator
crystal, and the radioactive sample (see Fig.~\ref{fig:schematic}). Similar
scintillator-camera detector configurations have been used for gamma- and
x-ray detection in previous studies \cite{4696615, 949368, Miller2014146,
1596911, 5401924, 5676230, nikl2006}. However, the present work differs in
that the entire camera is used as a ``single-channel'' detector, without
either energy resolution, imaging information, or photo-electron
amplification. The scientific camera could be used to detect gamma radiation
without the scintillation crystal \cite{peterson2011}. The thin silicon
camera architecture could be used as a kind of proportional detector.
However, our detector in this configuration is not as efficient as other
materials used previously \cite{2003A&A...411L.141L}.

A simplified schematic of our setup is shown in Fig.~\ref{fig:schematic}.  We
use the Andor Neo 5.5 CMOS camera. The camera's CMOS detector is cooled to
$-30^{\circ}$C, and the dark signal in the camera is less than 0.1
photo-electrons/pixel/second. A cylindrical NaI crystal, 25.4~mm in diameter
and 25.4~mm in length, is attached directly to the lens mount of the camera
via a light-tight housing (no focusing optics are used). The crystal is
encased in a sealed aluminum housing with a glass window on one end of the
cylinder allowing the scintillation photons to be detected by the camera. The
NaI crystal housing is temperature controlled to approximately $\pm$0.002
$^{\circ}$C. Although the NaI crystal scintillation efficiency is not overly
sensitive to temperature, the CMOS detector background level appears to
change when the crystal temperature changes.

In this setup, gamma radiation from the sample causes scintillation in the
crystal.  The scintillation light is detected by the CMOS chip below, and the
measured signal is the pixel gray-scale in each camera image. The data
reported here use the camera's high-sensitivity 16-bit digitization mode with
a rolling (electronic) shutter. In this mode, one detected photon results in
two ``counts'' on the camera. Our count rate data, measured in counts per
pixel per measurement time, is roughly analogous to average current
measurements in an ionization chamber.

\subsection{System stability}

\begin{figure}
\centerline{\includegraphics{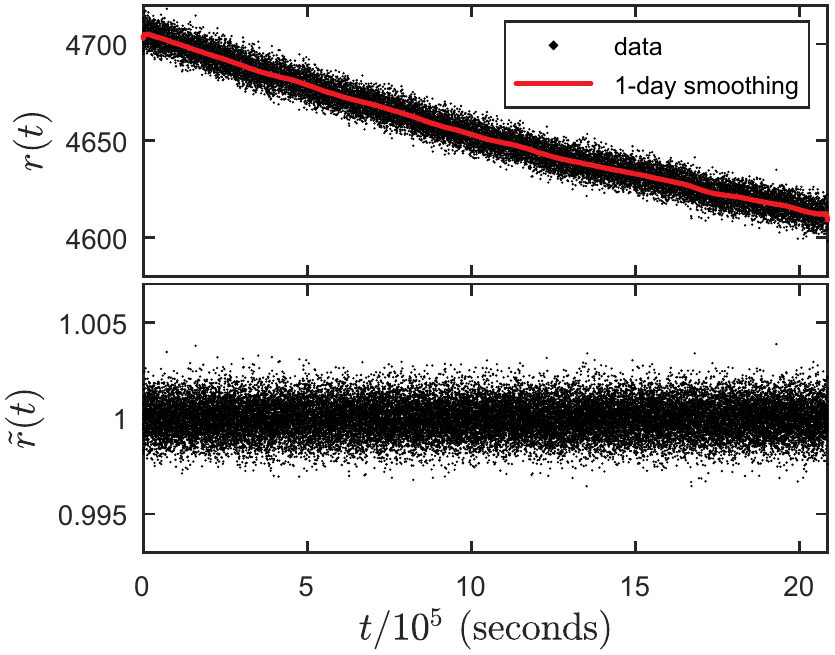}} \caption{\label{fig:stability}
(color online) Data used for stability analysis of the CMOS camera using a
\SI{10}{\micro Ci} Cs-137 gamma source. (a) The measured average number of
counts per pixel per measurement time, $r(t)$, plotted as black dots. Each
dot is the average signal over a 60 second interval. We observe a decline of
roughly 2\%, apparently due to a stability, dark signal, or amplifier gain
drift in the camera. Also shown is a line that represents the data smoothed
on a one day time scale. (b) The detrended count rate, $\tilde{r}(t)$, as
defined in the text. }
\bigskip

\centerline{\includegraphics{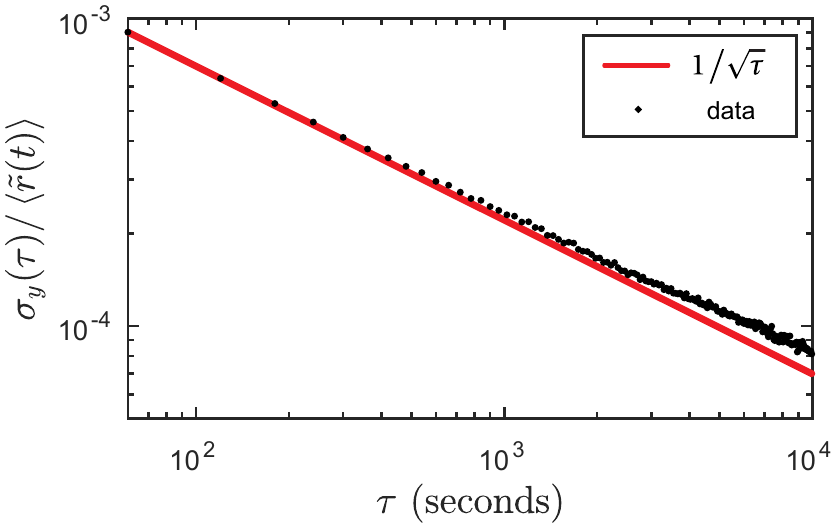}} \caption{\label{fig:stability2}
(color online) The The standard deviation of the detrended data in
Fig.~\ref{fig:stability}(b) as a function of the averaging time $\tau$.}
\end{figure}

Typical data from the experiment is shown in Fig.~\ref{fig:stability}(a). For
this data we place a \SI{10}{\micro Ci}  Cs-137 D-disk source in direct
contact with the NaI crystal housing so that the system approximates a 2$\pi$
detector. The NaI crystal is placed as close as conveniently possible to the
camera without any other optical elements. The camera integration time is set
to 60 seconds. For this test, we do not correct for the dark signal, and we
do not subtract off the small signal bias associated with the readout process
(fixed-pattern read noise). This is a direct test of the uncorrected
stability of the scintillator-plus-camera. Over the measurement time, the
Cs-137 activity is essentially constant. The camera pixels are binned into $8
\times 8$ ``super-pixels,'' and the average number of photo-electrons per
super-pixel, $r(t)$, is close to 5000 in 60 seconds.

In Fig.~\ref{fig:stability}(a), the measured count rate $r(t)$ drifts
downward by a few percent in over a period of about 24 days. To understand
the ultimate stability limit of the detection system, we separate variability
into two different time-scales---short term ``noise'' and long-term
``drift.'' To remove the long-term drift, we compare the individual data
points to the average of the surrounding day's data. This daily smoothing is
shown as the solid line in Fig.~\ref{fig:stability}(a). The corrected signal
$\tilde{r}(t)$, defined as the ratio of the raw data to the daily smoothing,
is plotted in Fig.~\ref{fig:stability}(b).

With the long-term drift removed, we analyze the remaining short-term noise
in Fig.~\ref{fig:stability}(b) by averaging the corrected data $\tilde{r}(t)$
over successively longer time intervals, and calculating the standard
deviation of these averages. The standard deviation of $\tilde{r}(t)$ as a
function of the averaging time $\tau$ is plotted in
Fig.~\ref{fig:stability2}. This shows that the statistical errors reach below
0.01\% in about 2 hours of data for the \SI{10}{\micro Ci} sample.  In
principle, this level of uncertainty could be reached faster by measuring a
sample with higher activity.

Understanding the long-term drift in Fig.~\ref{fig:stability}(a) is
critically important. It is not clear if this drift arises from the NaI
crystal or from the camera electronics. We have seen both positive and
negative drifts of similar fractional size while measuring NaI scintillation
from long-lived radio-isotopes and also while measuring a strongly attenuated
intensity-stabilized laser. This might suggest that the problem lies in the
camera. However, given our observations in Sec.~\ref{sec:time-response}, we
cannot conclusively eliminate NaI as a possible source of signal drift.

As we have shown in a previous publication, ratio measurement techniques can
remove detector drifts \cite{Bergeson2017}. If we consider alternating
measurements of two samples over a 4-hour time period, a drift of 2\% in a
week divides out to 0.04\% in four hours. For measurements intended to detect
drifts smaller than this, the source of the instability plotted in Fig.
\ref{fig:stability} will need to be reduced.

\subsection{System linearity}

\begin{figure}
\centerline{\includegraphics{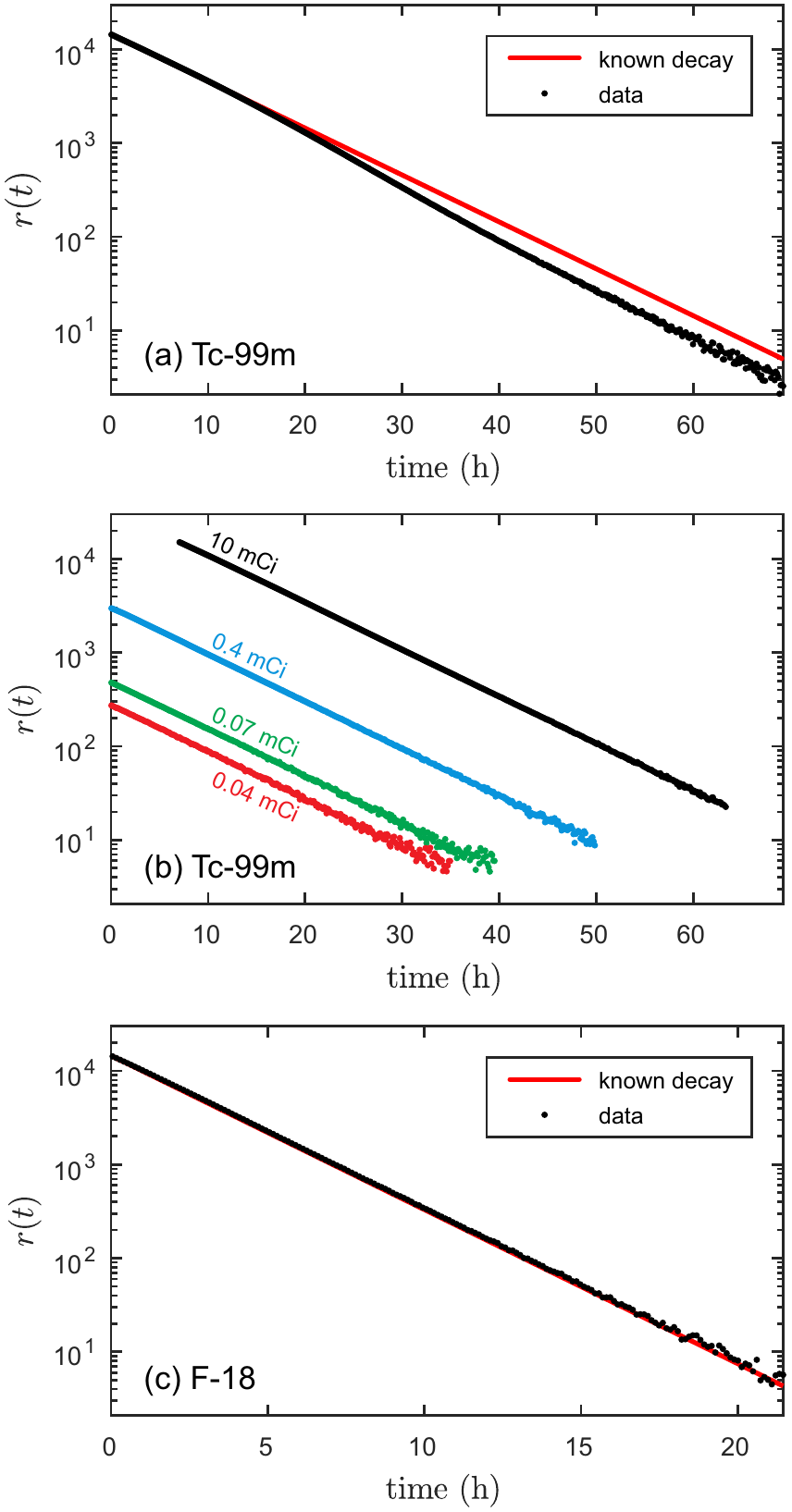}} \caption{\label{fig:linearity}
(color online) Linearity measurements and correction using short-lived
radio-isotopes. (a) Tc-99m. The black dots are the measured count rates and
the red line shows the expected count rate using the known half-life. The
departure from exponential decay is due to amplifier nonlinearity in the
camera. The initial sample activity is approximately 3 mCi. (b) Tc-99m data.
The black dots are the same data from panel (a), with the linearization
correction applied. The other data show the linearized measurements of lower
activity samples. The four fitted half-lives for this data are average to
6.00(2) days, close to the accepted value of 6.0067(10) days. (c) F-18 data.
The raw data is linearized using the Tc-99m correction. This produces a F-18
half life that is within 1\% of the accepted value. The black dots are the
linearized data, the red line is a fit with the known decay half-life. The
initial F-18 sample activity is about 3 mCi. }
\end{figure}

We characterize the linearity of the detection system by measuring the decay
of short-lived radio-isotopes over several half-lives. Measurements in
Tc-99m, with a half-life of $\tau_{1/2} = 6.0067\pm 0.0010 ~\mbox{days}$, and
F-18, with a half-life of $\tau_{1/2}^{ } = 1.82890 \pm 0.00023~\mbox{days}$
\cite{nuclideDotOrg}, are shown in Fig.~\ref{fig:linearity}. These samples
contain the radio-isotopes in a liquid form as a pertechnate for Tc-99m or
fludeoxyglucose for F-18. These samples are held in a sealed plastic syringe
above the NaI crystal. The crystal subtends a solid angle of 1.0~sr. In this
configuration, a 10~mCi sample exposes the crystal to $3\times 10^{7}$
gamma's per second (30~MBq).

The Tc-99m and F-18 scintillation data is recorded using the camera's 16-bit
image mode. In this mode, the camera uses a dual-amplifier configuration in
order to achieve the maximum dynamic range. One amplifier is optimized for
large signals, and the other is optimized for low signals. These amplifiers
are not perfectly matched, causing the signal to be slightly ``s''-shaped
relative to the expected exponential decay. In addition, there may be
nonlinearities in the NaI crystal scintillation. However, we use the measured
Tc-99m itself to linearize the overall system response. We find that this
linearization can be successfully applied to measurements of other isotopes.
This linearization gives excellent results for measurements spanning more
than three orders of magnitude in activity level.

In Fig.~\ref{fig:linearity}(a), we show the measured decay signal (black
dots) and the expected decay signal (red line) for a 3 mCi Tc-99m source. We
correct for the amplifier response using the ratio of the measured signal vs.
time for this source to the known decay rate. In Fig \ref{fig:linearity}(b)
we apply this correction to four subsequent measurements of Tc-99m samples
with different activities (10 mCi, 0.4 mCi, 0.07 mCi, and 0.04 mCi). The
linearization is excellent, reproducing the known half-life with sub-percent
residuals. We also measure the decay of a 3 mCi F-18 source and apply the
same linearity correction we used for the Tc-99m sample. The result is shown
in Fig.~\ref{fig:linearity}(c). The fitted lifetime matches the known
lifetime with an error of less than 1\%. As an additional check of the system
linearity, we measure the signal from several I-125 seeds. This isotope is a
low-energy gamma emitter with a half life of 59.388(28) days
\cite{nuclideDotOrg}. We place different activity samples in front of the NaI
crystal and measure the system response. The ratio of the measured response
to the known source activity matches the linearity function obtained from the
Tc-99m measurements to within the experimental uncertainties.

\subsection{Hour-long time-response of NaI}

\label{sec:time-response}

Finally, we turn to the time-dependence of the NaI scintillation when
measuring high activity samples. It is important to note that our detection
system is immune to pulse pile-up systematics. This is demonstrated in Fig.
\ref{fig:linearity}, where samples ranging from 0.04 to 10 mCi can be
measured without using any dead time correction.

\begin{figure}
\centerline{\includegraphics{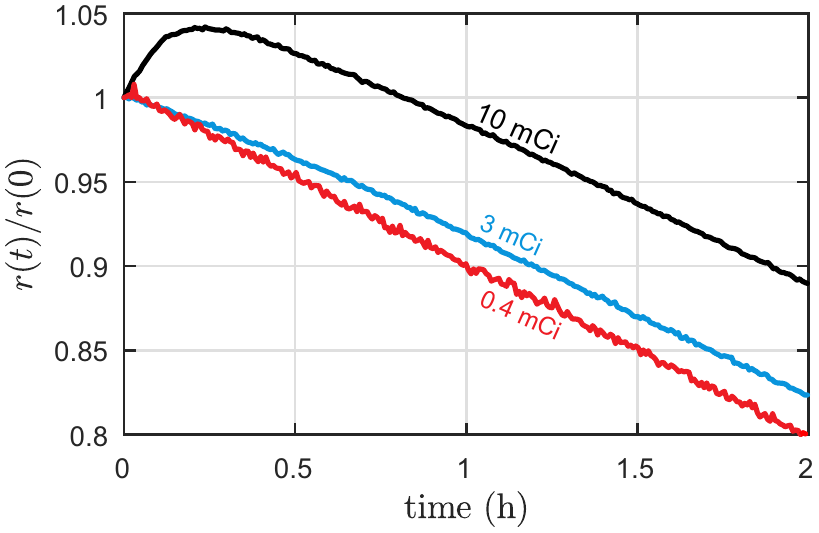}} \caption{\label{fig:time} (color
online) Measurements of three different Tc-99m samples for the first two
hours. The three measurements have been normalized so that the data overlap
at time zero, when the sample is first placed on the detector. The 0.3 mCi
source closely approximates the expected Tc-99m decay. The time-dependent
intensity non-linearity increases with sample activity, and more time is
required before the data approaches the expected exponential decay. Using
laboratory light sources, we verify that this behavior is not due to the
camera, but rather arises from the NaI crystal itself. Note that the early
time data leading to this effect was not included in the linearity
measurements of Fig.~\ref{fig:linearity}. }
\end{figure}

Data in Fig.~\ref{fig:time} show measurement data $r(t)$ of the first two
hours of three different Tc-99m samples, normalized by the initial signal
$r(0)$ measured just after the sample is placed on the detector. This data is
taken from the same data sets used in Fig.~\ref{fig:linearity}. However, for
that plot, the first few hours of nonlinear behavior was not used.

Figure~\ref{fig:time} shows a nonlinear response of the NaI scintillation
which increases with increasing activity levels. For samples with higher
activity rates, the NaI crystal takes longer to reach its steady state. For
the samples used in this study, this can take as long as a few hours.
However, when steady state is reached, the analysis in Fig.
\ref{fig:linearity} suggests that the decay of the short-lived Tc-99m and
F-18 isotopes can be accurately measured. There is some indication in our
data that NaI also takes some time to become completely dark after a
high-activity sample is removed from proximity to the crystal.

This short-term instability for NaI is inconvenient. It might be possible to
overcome the long-term stability issues using ratio measurements
\cite{Bergeson2017}. This would require excellent statistics for measuring
high activity samples on the hour-long time scale. Unfortunately, this is
precisely the time scale for short-term drifts in NaI response.

\section{Conclusion}

We report characteristics of a gamma-radiation detector based on a NaI
crystal and a scientific camera. The system is capable of measuring high
activity levels without dead time correction. With proper linearization, this
detector accurately measures short-lived radioisotopes over three decades of
dynamic range. We observe short-term changes in the NaI scintillation level
at high activity. We also observe long-term instabilities of a few percent
that may attributable to the camera, the NaI crystal, or both. These drift
characteristics must be considered for high precision measurements of
seasonal fluctuations in the nuclear decay rates.

Future work will need to isolate the source of drift seen in Fig.
\ref{fig:stability}. The camera could be tested in a better
temperature-controlled environment using an intensity-controlled light source
of verified stability. Future work should also study the response linearity
of other suitable scintillation materials. The analysis in
Fig.~\ref{fig:stability2} suggests that it should be possible to reach a
fractional accuracy of 0.0001\% in 7 days for a 1 mCi sample, assuming that
all sources of drift have been corrected. The camera+scintillator system is
capable of measuring even higher activity sources. However, it will require a
more linear scintillator than NaI.


\begin{thebibliography}{10}

\bibitem{nuclideDotOrg}
{Laboratoire Henri Bacquerel, Recommended Values}.
\newblock \url{http://www.nucleide.org/DDEP_WG/DDEPdata.htm}, 2016.
\newblock Accessed: 2016-09-23.

\bibitem{Bergeson2017}
S.~Bergeson, J.~Peatross, and M.~Ware.
\newblock Precision long-term measurements of beta-decay-rate ratios in a
  controlled environment.
\newblock {\em Physics Letters B}, 767:171 -- 176, 2017.

\bibitem{4696615}
S.~Caccia, G.~Bertuccio, D.~Maiocchi, et~al.
\newblock A mixed-signal spectroscopic-grade and high-functionality cmos
  readout cell for semiconductor x-$\gamma$ ray pixel detectors.
\newblock {\em IEEE Transactions on Nuclear Science}, 55(5):2721--2726, Oct
  2008.

\bibitem{Jenkins200942}
J.~H. Jenkins, E.~Fischbach, J.~B. Buncher, et~al.
\newblock Evidence of correlations between nuclear decay rates and
  {Earth–Sun} distance.
\newblock {\em Astroparticle Physics}, 32(1):42 -- 46, 2009.

\bibitem{2003A&A...411L.141L}
F.~{Lebrun}, J.~P. {Leray}, P.~{Lavocat}, et~al.
\newblock {ISGRI: The INTEGRAL Soft Gamma-Ray Imager}.
\newblock {\em Astron. Astrophys.}, 411:L141--L148, Nov. 2003.

\bibitem{949368}
D.~Meier, A.~Czermak, P.~Jalocha, et~al.
\newblock Silicon detector for a compton camera in nuclear medical imaging.
\newblock In {\em 2000 IEEE Nuclear Science Symposium. Conference Record (Cat.
  No.00CH37149)}, volume~3, pages 22/6--2210 vol.3, 2000.

\bibitem{5676230}
C.~M. Michail, V.~A. Spyropoulou, G.~P. Fountos, et~al.
\newblock Experimental and theoretical evaluation of a high resolution cmos
  based detector under x-ray imaging conditions.
\newblock {\em IEEE Transactions on Nuclear Science}, 58(1):314--322, Feb 2011.

\bibitem{5401924}
B.~W. Miller, L.~R. Furenlid, S.~K. Moore, et~al.
\newblock System integration of fastspect iii, a dedicated spect rodent-brain
  imager based on bazookaspect detector technology.
\newblock In {\em 2009 IEEE Nuclear Science Symposium Conference Record
  (NSS/MIC)}, pages 4004--4008, Oct 2009.

\bibitem{Miller2014146}
B.~W. Miller, S.~J. Gregory, E.~S. Fuller, et~al.
\newblock The iqid camera: An ionizing-radiation quantum imaging detector.
\newblock {\em Nuclear Instruments and Methods in Physics Research Section A:
  Accelerators, Spectrometers, Detectors and Associated Equipment}, 767:146 --
  152, 2014.

\bibitem{Nahle20158}
O.~N{\"{a}}hle and K.~Kossert.
\newblock Comment on “comparative study of beta-decay data for eight nuclides
  measured at the physikalisch-technische bundesanstalt” [astropart. phys. 59
  (2014) 47–58].
\newblock {\em Astroparticle Physics}, 66:8 -- 10, 2015.

\bibitem{nikl2006}
M.~Nikl.
\newblock Scintillation detectors for x-rays.
\newblock {\em Measurement Science and Technology}, 17, 2006.

\bibitem{peterson2011}
T.~E. Peterson and L.~R. Furenlid.
\newblock {SPECT} detectors: the anger camera and beyond.
\newblock {\em Phys. Med. Biol.}, 56, 2011.

\bibitem{Pomme2016281}
S.~Pomm\'{e}, H.~Stroh, J.~Paepen, et~al.
\newblock Evidence against solar influence on nuclear decay constants.
\newblock {\em Physics Letters B}, 761:281 -- 286, 2016.

\bibitem{Schrader2016202}
H.~Schrader.
\newblock Seasonal variations of decay rate measurement data and their
  interpretation.
\newblock {\em Applied Radiation and Isotopes}, 114:202 -- 213, 2016.

\bibitem{1596911}
C.~Stapels, W.~G. Lawrence, J.~Christian, et~al.
\newblock Solid-state photomultiplier in cmos technology for gamma-ray
  detection and imaging applications.
\newblock In {\em IEEE Nuclear Science Symposium Conference Record, 2005},
  volume~5, pages 2775--2779, Oct 2005.

\bibitem{ware2015}
M.~J. Ware, S.~D. Bergeson, J.~E. Ellsworth, et~al.
\newblock Instrument for precision long-term beta-decay rate measurements.
\newblock {\em Review of Scientific Instruments}, 86(7), 2015.

\end{thebibliography}

\end{document}